\documentclass[apjl]{emulateapj}

\usepackage{rotating}
\usepackage{graphicx}
\usepackage{epstopdf}
\usepackage{amsmath}

\newcommand{\msun}{~M$_\odot$}
\newcommand{\rsun}{~R$_\odot$}

\newcommand{\mearth}{~M$_\earth$}

\newcommand{\mjup}{~M$_{\rm J}$}
\newcommand{\mjupyr}{~M$_{\rm J}$~yr$^{-1}$}

\newcommand{\gcmcube}{~g~cm$^{-3}$}
\begin{document}

\title{On the survival of brown dwarfs and planets engulfed by their giant host star}
\author{Jean-Claude Passy$^{1,2}$, Mordecai-Mark Mac Low$^1$, Orsola De Marco$^3$}

\altaffiltext{1}{Department of Astrophysics, American Museum of Natural History, New York, NY, USA}
\altaffiltext{2}{Department of Physics and Astronomy, University of Victoria, Victoria, BC, Canada}
\altaffiltext{3}{Department of Physics and Astronomy, Macquarie University, Sydney, NSW, Australia}

\begin{abstract}
The recent discovery of two Earth-mass planets in close orbits around an evolved star has raised 
questions as to whether substellar companions can survive encounters with their host stars. 
We consider whether these companions could have been stripped of significant amounts of mass 
during the phase when they orbited through the dense inner envelopes of the giant.  
We apply the criterion derived by Murray et al. for disruption of gravitationally bound objects by ram pressure, 
to determine whether mass loss may have played a role in the histories of these 
and other recently discovered low-mass companions to evolved stars. We find that the brown dwarf 
and Jovian mass objects circling WD 0137-349, SDSS J08205+0008, and HIP 13044 are most unlikely 
to have lost significant mass during the common envelope phase. However, the Earth-mass planets 
found around KIC 05807616 could well be the remnant of one or two Jovian mass planets 
that lost extensive mass during the common envelope phase.
\end{abstract}


\section{Introduction}
\label{sec:intro}

The question of survivability of planets during post-main-sequence evolution of the host star 
has attracted some interest over the past years 
\citep{2007ApJ...661.1192V, 2010MNRAS.408..631N, 2012arXiv1208.2276S}. 
Recently, many substellar companions have  been discovered 
in close orbits around evolved stars (Table~\ref{tab:systems}). 
\cite{Maxted2006} detected a 0.053\msun \ brown dwarf orbiting the 0.39\msun \ white dwarf 
WD 0137-349 with a 1.9-hour period. A likely brown dwarf was discovered in a 2.3-hour orbit
around the subdwarf B star SDSS J08205+0008 \citep{GeierEtAl2011b}.
 \cite{SetiawanEtAl2010} discovered a Jupiter-mass object orbiting the red horizontal branch star 
 HIP 13044 with a 16.2-day period. Finally, \cite{CharpinetEtAl2011} reported the detection of 
 two nearly Earth-sized planets orbiting the subdwarf B star KIC 05807616 
 with orbital periods of 5.8 and 8.2 hours. The existence of these systems for which the radius 
 of the precursor of the primary was larger than today's orbital separation, suggests that 
 they have gone through a common envelope (CE) interaction. 
 During this phase, the companion is engulfed by its host star's envelope 
 when the latter was a giant star \citep{Paczynski1976}. 

Hydrodynamics simulations of this evolutionary phase have been performed 
with different numerical techniques \citep{Sandquist1998, PassyEtAl2012, RickerTaam2012} 
but questions persist. It is still unclear how the envelope of the donor star gets ejected, 
and why the final separations obtained by simulations are larger than those observed 
in post-CE systems \citep{PassyEtAl2012}. The resolution reached in these numerical models 
currently does not suffice to study the complete evolution for a substellar companion 
that may spiral down to very close to the giant's core. 

An important question is therefore whether substellar companions can survive engulfment 
by the giant's envelope without being totally disrupted. A system with a very low mass companion 
has a very small orbital energy budget, so we may also plausibly conjecture that a CE in such a system 
would more likely fail and result in a merger, thus destroying the companion. 
Is it therefore possible that substellar companions started off as more massive objects, 
which were then partially disrupted during their engulfment to become the objects we see today? 
This scenario was actually suggested by \cite{CharpinetEtAl2011} to explain 
the existence of planets around KIC 05807616. 

\citet{MurrayEtAl1993} studied the disruption of gravitationally bound objects 
by direct action of ram pressure in the context of star formation and galaxy evolution. 
However the physics of disruption is scale free, so their results also apply to other astrophysical questions 
such as the gas stripping occurring in clustered galaxies \citep{Nulsen1982, MoriBurkert2000} or here, 
the disruption of hydrostatic companions captured by the envelope of a giant star. 
In this contribution we use their theory to determine whether, in the observed systems, 
the substellar companions could have lost a significant amount of mass during the in-spiral phase. 
We describe the formalism used in Section~\ref{sec:analysis} and present 
the results in Section~\ref{sec:results}. 
We provide a summary in Section~\ref{sec:summary}.

\section{Analysis}
\label{sec:analysis}

\cite{MurrayEtAl1993} calculated the conditions for a self-gravitating, hydrostatic, isothermal sphere 
moving in an ambient medium to be disrupted by ram pressure.  The disruption of an object 
by ram pressure can most simply be understood by the pancake model \citep{Zahnle1992,KleinEtAl1994}, 
in which transverse motions are driven by the increased pressure on the front surface of the object 
compared to the sides. \citet{FieldFerrara1995} showed that the actual disruption by Kelvin-Helmholtz 
and Rayleigh-Taylor instabilities with wavelengths comparable to the size of the object 
reproduces the scalings yielded by this simple model. \citet{MurrayEtAl1993} confirmed numerically 
that when self-gravity is strong enough to stabilize the Kelvin-Helmholtz instability, 
the sphere is stable against catastrophic disruption. This occurs when  the gravitational acceleration
\begin{equation}
	g \gtrsim g_{\rm crit} \equiv \frac{2\pi \rho_2 \rho_1 U^2}{R_2(\rho_2^2 - \rho_1^2)} \ \ ,
\label{eq:gcrit}
\end{equation} 
\noindent where $\rho_1$, $\rho_2$, $R_2$ and $U$ are the background density, and the density, 
radius and relative velocity of the companion, respectively. We assume the tidal forces exerted 
by the giant star on the companion to be negligible compared with the self-gravity 
of the companion -- we discuss this point in Section~\ref{subsubsec:Soker}. 
Assuming $D \equiv \rho_2/\rho_1 \gg 1$ and taking the gravitational acceleration 
$g \approx G M_2/R_2^2$, one obtains a critical mass under which disruption 
occurs (their equation~2.8):
\begin{align}
M_{\rm crit} \sim 9.4~\left(\frac{50}{D}\right)^2 \left(\frac{U}{100~{\rm km\,s}^{-1}}\right)^3 \nonumber \\
\times \left(\frac{10^{-4}~{\rm g\,cm}^{-3}}{\rho_1}\right)^{1/2}~{\rm M}_{\rm J}.
\label{eq:mcrit}
\end{align}
If the configuration is unstable ($M_2 \lesssim M_{\rm crit}$), the companion is disrupted 
in a crossing time as the characteristic timescale for destruction is equivalent 
to its internal dynamical timescale:
\begin{equation}
	\tau \sim 1.4~\left(\frac{R_2}{0.1~{\rm R}_\odot}\right)\left(\frac{100~{\rm km\,s}^{-1}}{U}\right)\left(\frac{D}{50}\right)^{1/2} ~{\rm hr.}
\label{eq:timescale}
\end{equation}

\noindent We approximate the relative velocity of the companion by the maximum orbital velocity 
reached during the in-spiral phase, i.e. the final orbital velocity:

\begin{equation}
	U \sim \sqrt{\frac{GM_c}{a}}
\label{eq:velocity}
\end{equation}

\noindent where $M_c \gg M_2$ is the mass of the primary's core, and $a$ is the orbital separation 
at the end of the CE phase. We substitute this relation in Equation~(\ref{eq:mcrit}), 
and obtain the critical background density above which the companion is unstable:

\begin{align}
	\rho_{1,\rm crit} \sim 1.4~10^{-3}~\left(\frac{a}{{\rm R}_\odot}\right) \left(\frac{{\rm M}_\odot}{M_c}\right) \left( \frac{M_2}{{\rm M}_{\rm J}}\right)^{2/3} \nonumber \\
	 \times \left( \frac{\rho_2}{1~{\rm g\,cm}^{-3}}\right)^{4/3}~{\rm g\,cm}^{-3}.
\label{eq:rhocrit}
\end{align}

\noindent One should emphasize that if the companion is unstable, the dominant wavelength acting to destroy it is comparable to the sphere radius. The destruction is therefore global and not quasi-static. In other words, the companion is destroyed catastrophically, not ablated by incremental mass-loss due to short wavelength surface instabilities. Thus stability cannot be reached again once disruption starts. 

If the companion is stable ($M_2 \gtrsim M_{\rm crit}$), such shorter wavelength instabilities can still occur and strip mass in a more  ablative manner, giving a quasi-static mass loss rate.  We compute the largest unstable wavelength from Equations (\ref{eq:gcrit}) and (\ref{eq:velocity}), again assuming $\rho_2 \gg \rho_1$, and substituting $\lambda_{\rm max}$ for $R_2$. We find

\begin{align}
	\lambda_{\rm max} = 7.8~10^{-2}\left(\frac{M_c}{{\rm M}_\odot}\right) \left(\frac{{\rm R}_\odot}{a}\right)  \left( \frac{1~{\rm g\,cm}^{-3}}{\rho_2}\right)^{5/3} \nonumber \\
	\times \left(\frac{\rho_1}{10^{-4}~{\rm g\,cm}^{-3}}\right)   \left(\frac{{\rm M}_{\rm J}}{M_2}\right)^{1/3}~{\rm R}_{\rm J}.
\end{align} 

\noindent \cite{MurrayEtAl1993} suggests that the corresponding mass loss rate can then be approximated to that occurring in laminar viscous flows as described by their Equation~3.5:

\begin{align}
	\dot{M}_{\rm v} \sim 11 \left(\frac{U}{100~{\rm km\,s}^{-1}}\right) \left(\frac{\rho_1}{10^{-4}~{\rm g\,cm}^{-3}}\right) \nonumber \\ 
	\times   \left(\frac{R_2}{0.1~{\rm R}_\odot}\right)^2 \left(\frac{\lambda_{\rm max}}{R_2}\right)~{\rm M}_{\rm J}\,{\rm yr}^{-1}
	\label{eq:mdot_stable}
\end{align} 

\noindent where we have assumed the relative velocity to be of the order of the sound speed 
of the background. Equation~(\ref{eq:mdot_stable}) should be used with caution as it is valid only 
for Reynolds numbers $Re \lesssim 30$ \citep{Nulsen1982}. In our case, the Reynolds number is 
$Re = 2 R_2 U / \nu$, where $\nu$ is the molecular viscosity. Viscosity in stellar interiors 
is far too small to allow such low Reynolds numbers on these length scales.

However, such low effective Reynolds numbers might be reached through the turbulent viscosity. 
Since the largest possible size of the eddies is $\lambda_{\rm max}$ and their largest velocity 
is the sound speed of the ambient medium $c_S$, one can use a formalism 
like that of \citet{ShakuraSunyaev1973} and parametrize the turbulent viscosity as:

\begin{equation}
	\nu_{\rm t} = \alpha c_S \lambda_{\rm max}
\end{equation}

\noindent where $\alpha \lesssim 1$ is a free parameter. The effective Reynolds number is thus:

\begin{equation}
	Re_{\rm t} \sim \frac{2 R_2}{\alpha \lambda_{\rm max}}
\end{equation}

\noindent where we have still assumed that $U \sim c_S$. Unfortunately, no accepted value of $\alpha$ 
has been derived for our specific case. We can only calculate the minimum turbulent Reynolds number
$Re_{\rm t, min} = Re_{\rm t}(\alpha = 1)$ and see how this number compares 
to the range required to apply Equation~(\ref{eq:mdot_stable}). 

Another approach would be to calculate how much work from the drag acts on the companion 
during its in-spiral, and compare it to the companion's gravitational binding energy $E_{\rm bin} \sim GM_2^2/R_2$. 
Assuming for the cross section $S_2 = \pi R_2^2$ and a drag coefficient $C_D$, 
the drag force is $F_D = \frac{1}{2} C_D \rho_1 U^2 S_2$. 
The smaller the companion, the longer the in-spiral phase lasts: for example,
the 0.9\msun \ and 0.1\msun \ companions complete about 20 and 40 orbits during this phase, respectively \citep{DeMarcoEtAl2012,PassyEtAl2012}.
 We therefore assume that our companions complete at least 50 orbits, and slice the giant's envelope 
 down to the currently observed separation of the system, in $N_{\rm orbits} = 50$ equally-spaced radii 
 $\bar{R_i}$ corresponding to an average density $\bar{\rho_i}$. 
 Assuming $C_D = 1$, the work provided by drag forces is thus:

\begin{equation}
	W_{\rm drag} = \sum_{i \leq N_{\rm orbits}}  \frac{1}{2} \bar{\rho_i} U_i^2 S_2 \times 2\pi \bar{R_i}.
	\label{eq:drag}
\end{equation}

\noindent where $U_i = \sqrt{GM_i/\bar{R_i}}$ and $M_i$ is the mass enclosed within $\bar{R_i}$. 
We will see in Section~\ref{subsec:Charpinet} that the calculation of $W_{\rm drag}$ does not depend 
strongly on the number of assumed orbits. If $W_{\rm drag} \gtrsim E_{\rm bin}$ \
we conclude that the companion could have been 
destroyed if the drag energy can be coupled to the interior of the companion efficiently, 
even if it is stable according to Equation~(\ref{eq:gcrit}).

\section{Results}
\label{sec:results}

In this section we study each system reported in Table~\ref{tab:systems} separately and investigate 
whether the substellar components are less massive today than they were prior to the CE evolution.

\subsection{WD 0137--349}
\label{subsec:maxted}

\begin{figure}[h!]
	\begin{center}
		\includegraphics[scale=0.42]{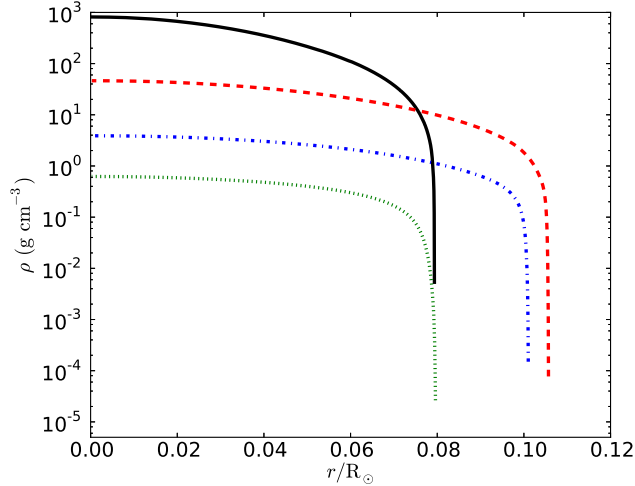}
	\caption{Density profiles for the 0.053\msun \ (solid black), 10\mjup \ (dashed red), 1\mjup \ (dash-dotted blue) and 0.1\mjup \ (dotted green) models.
	\label{fig:profiles}
	}
	\end{center}
\end{figure}

Using the {\it MESA} stellar evolution code  \citep{PaxtonEtAl2011}, we compute a model for the 0.053\msun \ brown dwarf companion to this white dwarf and obtain a radius 
$R_2 = 0.079$\rsun \ (Figure~\ref{fig:profiles}).
The mean density of the brown dwarf is $\rho_2 = 150$\gcmcube \ which, 
substituted in Equation~(\ref{eq:rhocrit}), yields a critical density $\rho_{1,\rm crit} \approx 27$\gcmcube.

\begin{figure}[h!]
	\begin{center}
		\includegraphics[scale=0.4]{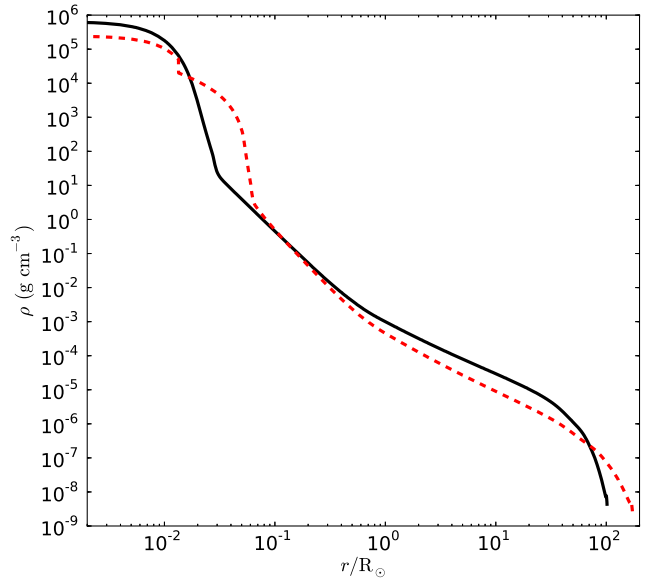}
	\caption{Density profile for the 0.89\msun \ red giant star progenitor of WD 0137--349 (solid black) and a 0.76\msun \ star at the tip of the red giant branch (dotted red).}
	\label{fig:progenitor}
	
	\end{center}
\end{figure}

We compare this critical density to the density profile of the progenitor and see whether the companion 
will plunge deep enough into the primary's envelope to encounter a density high enough to destroy it. 
The most likely progenitor of the white dwarf is a nearly 1\msun \ main sequence star. 
We evolve such a model with {\it MESA} up to the red giant branch until the core mass reaches 
the observed white dwarf mass. At that time, the radius of the star was 100\rsun \ and its total mass 
was 0.89\msun \ due to mass loss. The density profile of this red giant star is displayed 
in Figure~\ref{fig:progenitor}. The current orbital separation of the system is $a=0.65$\rsun \ 
which means that the companion has reached a layer of the primary 
where the density $\rho_{1, \rm max} \sim 2.1 \times 10^{-3}$\gcmcube. 

At this point, we should emphasize that we have calculated the maximum density 
encountered by the companion using the {\it initial} profile of the giant primary. 
Hydrodynamical simulations suggest that in the deep interior, density at a given coordinate 
decreases with time \citep[Figure~12] {PassyEtAl2012} although the current resolution 
does not allow them to give a definitive answer at such small radii. 
Therefore, $\rho_{1, \rm max}$ should be considered an upper limit.

$\rho_{1, \rm max}$ is four orders of magnitude smaller than the critical density required for instability 
so we conclude that the companion is stable. Moreover, the mass loss rate due to viscous stripping 
$\sim 3.8\times 10^{-2}$\mjupyr \ only. Since the dynamical phase of a CE lasts only about one year \citep{Sandquist1998, PassyEtAl2012}, 
the companion has not lost any significant amount of mass during its in-spiral phase.

\subsection{SDSS J08205+0008}
SDSS J08205+0008 is very similar to WD 0137--349 so we also conclude that the brown dwarf -- 
whose estimated mass ranges from 0.045 to 0.068\msun \ --  has not been affected during the plunge-in phase.

\subsection{HIP 13044}

The minimum mass of the planet orbiting HIP 13044 is estimated to be 1.25\mjup \ \citep{SetiawanEtAl2010}.
This system is intriguing, as a plausible formation channel for this system has yet to be found. 
Indeed, the orbital separation ($\sim 25$\rsun) is much larger than the separations obtained 
from numerical simulations or observations of post-CE systems \citep{AlphaPaper2011, PassyEtAl2012}. 
Also, successful CE interactions usually lead to the formation of blue horizontal branch stars 
with very thin hydrogen-envelopes \citep{HanEtAl2002}. 

\cite{BearEtAl2011} therefore suggested that the helium flash occurring in a 0.9\msun \ metal-poor red giant star
could ignite hydrogen burning in a thick region at the bottom of the envelope. 
This deposition of energy would eventually lead to a dramatic inflation lasting about 100\,yr, during which the stellar radius could reach 700\rsun. The planet would then enter a ``half-failed, half-successful'' CE 
during which the orbital separation would shrink down to approximately 25\rsun, only 0.1\msun \ would be lost 
by the system and the primary would evolve to the red part of the horizontal branch. 

This theory has several problems. First, there is no compelling evidence suggesting that 
red giant branch stars might suffer such a large and long-lasting expansion after the helium core flash. 
On the contrary simulations performed with stellar evolution codes such as {\it MESA}, show 
that red giant stars shrink after helium has been ignited explosively. 
Secondly, the duration of this suggested post-helium-flash expansion is larger than the duration of a CE phase. 
Therefore, the in-spiral of the companion could not have been stopped because of the primary 
contracting below its Roche lobe on a short timescale. Thus, the entire envelope must have been lifted up, 
mostly in the equatorial plane \citep{PassyEtAl2012, RickerTaam2012}, which raises thus the question of 
how a red horizontal branch star could eventually form. Finally, the planetary companion would strongly 
interact with falling back material through gravitational drag. In that regard, the large orbital separation observed 
is still perplexing. For all these reasons we believe that the scenario 
suggested by \cite{BearEtAl2011} is unlikely.

A careful investigation of the progenitor of HIP 13044 seems quite complex and beyond the purposes 
of this paper. Consequently we simply assume that the companion went through a CE interaction 
with an unknown progenitor. For the latter, we will consider the same profile as in Section~\ref{subsec:maxted} 
in order to give an indicative answer on the potential survival of the companion.

For the companion, we consider a 1\mjup \ model (Figure~\ref{fig:profiles}), with a radius 
$R_2 = 0.101$\rsun \ and a mean density $\rho_2 = 1.31$\gcmcube. We find the critical density for 
stability $\rho_{1,\rm crit} \sim 6.2\times 10^{-2}$\gcmcube. The density in the progenitor's profile 
corresponding to the observed separations between the subdwarf and 
the planet $\rho_{1, \rm max} \sim 10^{-5}$\gcmcube, almost 4 orders of magnitude smaller 
than $\rho_{1,\rm crit}$. We thus conclude that the companion is stable. 
The viscous mass loss rate is also negligible ($\dot{M}_{\rm v} \sim 1.4\times 10^{-4}$\mjupyr) 
so no mass will be ablated from the companion during the plunge-in phase.

\subsection{KIC 05807616}
\label{subsec:Charpinet}

The most intriguing system is certainly KIC 05807616, which has two nearly Earth-mass planets 
orbiting a subdwarf B star with an orbital separation of about 1.5\rsun. In order to constrain 
the mass of these planets prior to their engulfment, we use the same method as previously, 
with four different models ranging from 10\mjup \ down to 1\mearth \ (Figure~\ref{fig:progenitor}). 
For the Earth-sized model, we use the accepted values for the Earth \citep[see, e.g.,][]{Cox2000} 
as listed in Table~\ref{tab:planets}. We again assume the primary used in Section~\ref{subsec:maxted} 
for which we obtain $\rho_{1, \rm max} \approx 10^{-3}$\gcmcube. 

The results are summarized in Table~\ref{tab:planets}. One can see that the critical density required to 
catastrophically destroy an Earth-sized companion is about as large as the maximum density  
$\rho_{1, \rm max}$. In this regard, the 1\mearth \ companion could still be stable. 
However $W_{\rm drag} \sim 10^{41}$~ergs, which is more than an order of magnitude larger than 
the gravitational binding energy of the companion (Table~\ref{tab:planets}). $W_{\rm drag}$ does not 
strongly depend on the number of assumed orbits since changing this number to 20 or 100 
changes $W_{\rm drag}$ by a factor 2--3 only. Since the work provided by drag forces 
exceeds the binding energy of the companion, we conclude that the Earth-sized companion 
may not survive the engulfment, if the energy can be effectively coupled to the planetary material.

From the same considerations, a 10\mjup \ companion is too massive to be affected 
in any way during its in-spiral phase (Table~\ref{tab:planets}). For a 1\mjup \ companion, 
$\rho_{1, \rm max}$ exceeds $\rho_{1,\rm crit}$ by almost an order of magnitude, so the secondary 
could be stable. However,
 $E_{\rm bin}$ exceeds $W_{\rm drag}$ by only a factor of 3, which suggests that a significant amount 
of mass could be removed from the surface of the companion. A 0.1\mjup \ companion 
has a critical density an order of magnitude smaller than $\rho_{1, \rm max}$. 
Moreover, $W_{\rm drag}$ is about twenty times larger than the binding energy of the companion.  
Together, these suggest that an 0.1\mjup \ object probably would be destroyed. 
Using a more evolved model with a higher core mass $M_c = 0.47$\msun \ 
and $N_{\rm orbits} = 100$ (the stellar radius $\sim 175$\rsun)
yields similar results (Figure~\ref{fig:progenitor}, Table~\ref{tab:planets})
We thus conclude that the progenitor of the planets likely had a mass of a few \mjup, 
supporting the suggestion made by\cite{CharpinetEtAl2011}.

\subsubsection{Discussion}
\label{subsubsec:Soker}

A legitimate question is whether mass accretion by the substellar companion is relevant. 
From a Bondi-Hoyle analysis \citep[see, e.g.,][equation~31]{Edgar2004} the accretion rate 
onto the 1\mjup \ object is only

\begin{equation}
	\dot{M}_{\rm acc} \sim \frac{\pi G^2 M_2^2 \rho_{1, \rm max}}{\sqrt{2} U^3} = 3.7 \times 10^{-2}~{\rm M}_{\rm J}/{\rm yr}.
	\label{eq:BH}
\end{equation}

\noindent We therefore conclude that mass accretion is negligible on the timescale of the CE interaction. 
\cite{RickerTaam2008} also reached this conclusion in the case of more massive secondaries. 
They further argued that the mass accretion rate calculated from their simulations is even smaller 
than the one given by Equation~(\ref{eq:BH}), mainly because tidal effects dominate the structure of the flow.

\cite{BearSoker2012} propose a slightly different scenario for the formation of the two planets 
around KIC 05807616. Instead of considering the dynamical effects occurring during the CE phase, 
they suggest that a massive giant planet was tidally destroyed after the end of the in-spiral phase 
and that the two planets are remnants of the disrupted metallic core. 
Using their equation~1 with $C_{\rm tide}  =1$, the values given by \cite{CharpinetEtAl2011}, 
and our planetary models (Table~\ref{tab:planets}), the tidal radius at which destruction occurs is 
$R_t = 0.39$ and $0.80$\rsun \ for the 10 and 1\mjup \ objects, respectively. 
These qualitative values are still smaller than the observed separations of 
the planets (see Table~\ref{tab:systems}) so it is difficult to conclude whether these objects 
would be tidally destroyed. They also suggest that evaporation of the companion due to heating 
from the envelope might be important. They estimate the mass of the planet must be $\gtrsim 5$\mjup \ 
in order to survive evaporation. The evaporation timescale depends on the efficiency of heat transfer, 
which is sensitive to mixing between the planetary atmosphere and the stellar convective envelope. 
We cannot at present calculate this timescale with sufficient precision. We therefore do not know 
whether evaporation could dominate the dynamical effects we discuss in this study. 
Nevertheless, the mass of the progenitor suggested by \cite{BearSoker2012} is consistent with ours.

\section{Summary}
\label{sec:summary}

We have considered evolved stars found to have brown dwarfs or planets 
around them in close orbits, suggesting that these systems have gone through a CE interaction. 
Using the criterion developed by \citet{MurrayEtAl1993}, 
we have investigated the stability of these substellar companions against mass loss 
induced by Kelvin-Helmholtz and Rayleigh-Taylor instabilities. 

We have found that the substellar companions observed in WD 0137--349, SDSS J08205+0008 
and HIP 13044 are very unlikely to have been affected during the in-spiral inside their giant progenitors. 
Therefore, the masses of these objects currently observed are likely the same as prior to the CE phase. 
The two planets detected orbiting KIC 05807616, however, are not massive enough to have survived 
the engulfment unscathed. We have estimated the mass of the progenitor of these planets to be 
of order 1\mjup. The question that still has to be answered is how such low mass companions 
could have successfully unbound the entire envelope of the primary.

\section{Acknowledgments}
\label{sec:acks}
We acknowledge funding from NSF grant AST-0607111. We are thankful to Noam Soker, 
Alexander Hubbard, Phil Arras for their contributions that helped improve this manuscript, 
and Bill Paxton for making {\it MESA} publicly available. J-CP thanks Falk Herwig for his support.


\begin{deluxetable}{cccccc}
\tabletypesize{\scriptsize}
\tablewidth{0pt} 
\tablecaption{Orbital parameters.}
\tablehead{
	    \colhead{Name}               &
           \colhead{$M_{c}$~(M$_\odot$)}        &
           \colhead{$M_2$}      &
           \colhead{$P$}       &
           \colhead{$a$~(R$_\odot$)}   &
           \colhead{References}}
\startdata
WD 0137--349 & 0.39 & 55.6\mjup & 1.93~hours & 0.65 & \cite{Maxted2006} \\
SDSS J08205+0008 & 0.25-0.47 & 47.1 -- 71.2\mjup & 2.30~hours & 0.60 -- 0.72 & \cite{GeierEtAl2011b} \\
HIP 13044 & 0.8 & $\geq$1.25\mjup & 16.2 days & 24.95 & \cite{SetiawanEtAl2010} \\
KIC 05807616 & 0.496 & 0.440\mearth & 5.7625 hours & 1.290 & \cite{VanGrootelEtAl2010, CharpinetEtAl2011} \\
KIC 05807616 & 0.496 & 0.655\mearth & 8.2293 hours & 1.636 &  \cite{VanGrootelEtAl2010, CharpinetEtAl2011}
\enddata
\label{tab:systems}
\tablecomments{Reported are the stellar mass ($M_c$), the companion mass ($M_2$), the orbital period ($P$) and the orbital separation ($a$)}
\end{deluxetable}

\begin{deluxetable}{ccccccccc}
\tabletypesize{\scriptsize}
\tablewidth{0pt} 
\tablecaption{Parameters of the different companion models investigated for KIC 05807616.}
\tablehead{
	    \colhead{$M_2$} &
           \colhead{$R_2$~(R$_\odot$)} &
           \colhead{$\rho_2$~(g.cm$^{-3}$)} &
           \colhead{$\rho_{1,\rm crit}$~(g.cm$^{-3}$)} &
           \colhead{$\lambda_{\rm max} / R_2$} & 
           \colhead{$Re_{\rm t, min}$} &       
           \colhead{$\dot{M}_{\rm v}$~(M$_{\rm J}$\,yr$^{-1}$)} &          
           \colhead{$W_{\rm drag}$~(ergs)}  &
           \colhead{$E_{\rm bin}$~(ergs)}}
\startdata
10\mjup & 0.106 & 11.4 & 0.5 & $2.0(-3)$ & $9.9(2)$ &$0.6$& $1.2(43)$& $3.3(45)$\\
1\mjup & 0.101 & 1.31 & $6.0(-3)$ & $1.7(-1)$& 12 & 45 & $1.1(43)$ & $3.4(43)$ \\
0.1\mjup & 0.080 & 0.27 & $1.5(-4)$ & -- & -- & -- & $7.0(42)$ & $4.3(41)$\\
1\mearth & $9.16(-3)$ & 5.52 & $8.8(-4)$ & -- & -- & -- & $1.1(41)$ & $3.7(39)$
\\
\\
10\mjup & 0.106 & 11.4 & 0.5 & $1.0(-3)$ & $2.0(3)$ &$0.15$& $5.3(42)$& $3.3(45)$\\
1\mjup & 0.101 & 1.31 & $6.0(-3)$ & $8.4(-2)$& 24 & 12 & $4.8(42)$ & $3.4(43)$ \\
0.1\mjup & 0.080 & 0.27 & $1.5(-4)$ & -- & -- & -- & $3.0(42)$ & $4.3(41)$\\
1\mearth & $9.16(-3)$ & 5.52 & $8.8(-4)$ & -- & -- & -- & $4.9(40)$ & $3.7(39)$
\enddata
\label{tab:planets}
\tablecomments{Values obtained with the 0.89\msun \ (top) and the 0.76\msun \ (bottom) models. Quantities relative to viscous stripping are not calculated when the object is found unstable.}
\end{deluxetable}

\end{document}